\documentclass[aps,pre,superscriptaddress,twocolumn]{revtex4}
\usepackage{graphicx}

\begin{document}

\title{Error-driven Global Transition in a Competitive Population on a Network}

\author{Sehyo Charley Choe}
\author{Neil F. Johnson}
\affiliation{Clarendon Laboratory, Physics Department, Oxford University, Oxford OX1 3PU, U.K.}

\author{Pak Ming Hui}
\affiliation{Department of Physics, The Chinese University of Hong Kong, Shatin, New Territories, Hong Kong}

%\date{\today}

\begin{abstract} We show, both analytically and numerically, that erroneous data transmission generates a global transition within a competitive population playing the Minority Game on a network. This transition, which resembles a phase transition, is driven by a `temporal symmetry breaking' in the global outcome series. The phase boundary, which is a function of the  network connectivity $p$ and the error probability $q$, is described quantitatively by the Crowd-Anticrowd theory.
 
\end{abstract}

\maketitle

The study of Complex Adaptive Systems is enabling Physics to expand its
boundaries into a range of non-traditional areas within the biological, informational and socio-economic communities. Since these areas are often rich in empirical data, they also offer Physics a new testing ground for theories of Complex Systems, and for non-equilibrium statistical mechanics in general \cite{examples}. It is widely recognized \cite{Casti} that Arthur's multi-agent El Farol Bar Problem (EFBP) \cite{Arthur1} embodies the interacting, many-body nature of many real-world Complex Systems. In particular, a binary representation of the EFBP, namely the Minority Game (MG) \cite{Challet1,NFJ1,TMG,netMG}, plays the role of a new `Ising Model' for theoretical physics.

Despite widespread interest among physicists in biological, informational and socio-economic networks \cite{nets}, researchers have only just started
considering the effect of such networks in the MG and EFBP \cite{netMG}. It
has so far been assumed that any information shared between agents is  always perfectly accurate. However, real-world Complex Systems do not operate at such levels of perfection. Furthermore, informational networks might be used by agents to spy and mislead rather than to benefit others. This raises the following question: what is the effect of networks within a
competitive population such as the MG, where the information transmitted is
corrupted? 

Here we show analytically and numerically, that erroneous data transmission generates an abrupt global transition within a competitive, networked population playing the Minority Game. This phase-like transition is driven by a `temporal symmetry breaking' in the global outcome series. The Crowd-Anticrowd theory, which accounts for the many-body (i.e. many-agent) correlations inherent in the system, provides a quantitative yet physically intuitive explanation of this phase transition.

Our model consists of $N$ objects or `agents' who repeatedly compete to be in a minority: for example, commuters striving to choose the least crowded of two routes. The agents can be any form of adaptive object, e.g. biological or mechanical, and our general setup has potential application to a wide range of problems in the biological, informational and social  sciences. The minority rule can also be generalized \cite{NFJ1}.  At each timestep $t$, each agent decides between action $+1$ meaning to choose option `1',  and action $-1$ meaning to choose option `0'. The winning (i.e. minority) outcome at each timestep is  `0' or `1'. Each agent decides his actions in light of (i) {\em global information} which takes the form of the history of the $m$ most recent global outcomes, and (ii) {\em local information} obtained via the cluster to which he is connected, if any. Such connections may be physically tangible (e.g. a telephone or Internet link, or biological structure) or physically intangible (e.g. a wireless communication channel, or biochemical pathway). Adaptation is introduced by
randomly assigning $S$ strategies to each agent. Each of the $2^{2^m}$ possible strategies is a bit-string of length $2^m$ defining an action ($+1$ or $-1$) for each of the $2^m$ possible global outcome histories
$\mu(t)$ \cite{Challet1,NFJ1}. For $m=2$ for example, there are  $2^{2^{m=2}}=16$ possible strategies and $2^2=4$ possible global outcome histories: $\mu(t)=00$, $01$, $10$ and $11$. Strategies which predicted the winning (losing) action at a given timestep are assigned (deducted) one point. 

Agents use the connections they have, if any, to gather information from other agents. For simplicity we assume a random network between agents, fixed at the beginning of the game. The connection between any two agents exists with a probability $p$, hence each agent is on average connected to $p(N-1)$ others. At a given timestep $t$, and with a given global history $\mu(t)$, each agent takes the action predicted by the highest-scoring strategy among his own {\em and} those of the agents to which he is connected. The parameter $q$ is the probability that an {\em error} arises in the information he gathers from his cluster. Alternatively, $q$ can be viewed as the weight an agent places on the information gathered from his cluster. For example, if the action of the best strategy in his cluster is $+1$, the agent records this as a $-1$ with probability $q$ (and vice versa for a best action $-1$). The information transmission has been corrupted with probability $q$.  Any agent with a higher-scoring strategy than those of his neighbors at a given timestep, is unaffected by this error -- the only source of stochasticity which might affect him is the standard coin-toss used to break any ties between his own strategies \cite{NFJ1}. In contrast to the agents' `on-site' stochastic strategy selection arising in the Thermal Minority Game (TMG) \cite{TMG,qNote}, the stochasticity associated with  $q$ in our game depends on the agents' connectivity. 

%%%%%%%%%%%%%%%%%%% Figure 1: Phase Portrait %%%%%%%%%%%%%%%%%%%%%%%%%
\begin{figure}
\scalebox{0.7}{\includegraphics{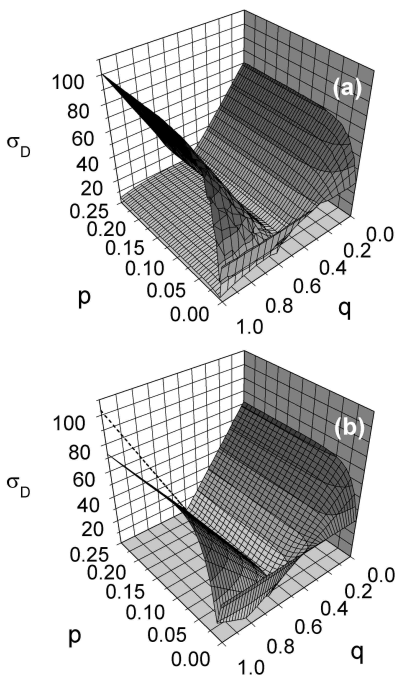}}
\caption{Fluctuation in excess demand, $\sigma_{D}$, as a function of the
error-probability $q$ and the network connectivity $p$. (a)  Numerical results averaged over 300 runs,  each with $10^{5}$ iterations.  (b) Analytic Crowd-Anticrowd theory. At high $q$, the two branches in (a) correspond to different dynamical attractor states, while the single branch in (b) represents an effective average (see text).  The dotted line in (b) at $p = 0.25$, illustrates the modified analytical results for the upper branch if one assumes some knowledge of this branch's global output series. Parameters: $m=1$, $S=2$, and $N=101$. }
\label{figure1}
\end{figure}
%%%%%%%%%%%%%%%%%%%%%%%%%%%%%%%%%%%%%%%%%%%%%%%%%%%%%%%%%%%%%%%%%%%%%%

%%%%%%%%%%%%%%%%%%%%%% Figure 2: q>0.5 Zoom %%%%%%%%%%%%%%%%%%%%%%%%%%
\begin{figure}
\scalebox{0.7}{\includegraphics{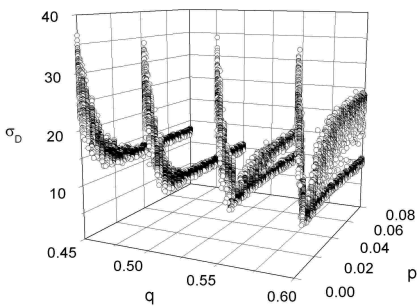}}
\caption{Numerical results for individual runs, showing the fluctuation in
excess demand, $\sigma_{D}$,  around $q=0.5$. Parameters as in Fig. 1.}
\label{figure2}
\end{figure}
%%%%%%%%%%%%%%%%%%%%%%%%%%%%%%%%%%%%%%%%%%%%%%%%%%%%%%%%%%%%%%%%%%%%%%

%%%%%%%%%%%%% Figure 3: Minimum Excess Demand Channel %%%%%%%%%%%%%%%%
\begin{figure}
\scalebox{0.7}{\includegraphics{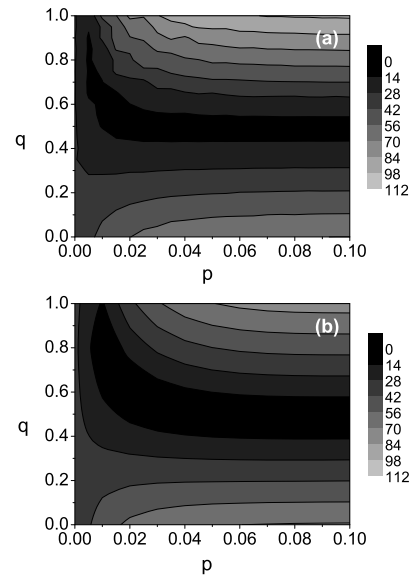}}
\caption{Contour map version of Figure 1. Contours correspond to a constant value of $\sigma_{D}$ as a function of the error-probability  $q$ and the network connectivity $p$. (a) Numerical results.  (b) Analytic Crowd-Anticrowd theory.  For clarity, only the upper branch of the numerical results is shown for the high-$q$ phase.}
\label{figure3}
\end{figure}
%%%%%%%%%%%%%%%%%%%%%%%%%%%%%%%%%%%%%%%%%%%%%%%%%%%%%%%%%%%%%%%%%%%%%%

We now investigate the effects of this {\em microscopic} connection-driven data error on the system's {\em macroscopic} dynamics. We shall build an analytic theory based on the Crowd-Anticrowd theory \cite{NFJ1} which incorporates the many-agent (many-body) correlations arising in the system's strategy space as a result of the dynamics in the history space.  To a very good approximation, we can replace the Full Strategy Space by a Reduced Strategy Space (RSS) \cite{Challet1} which provides a minimal basis set of strategies for the system \cite{NFJ1}. The appropriate choice for the RSS depends on the relative frequency of visits to the $2^m$ histories.  With all histories visited equally often, the RSS comprises a total of $2P=2.2^{m}$ strategies \cite{Challet1}.  As in previous MG work \cite{Challet1}, we will focus on the typical fluctuations in the excess demand $D(t)$ away from its optimal value as a function of $p$ and $q$.  The excess demand $D(t)$ at timestep $t$ is given by: 

\begin{equation}  
D(t)  =  n_{-1}(t) - n_{+1}(t) = \sum_{K=1}^{K=2P} a_{K} n_{K}(t)
\label{D}
\end{equation}

\noindent where $n_{+1}(t)$ ($n_{-1}(t)$) is the number of agents taking action $+1$ ($-1$); $a_{K}=\pm1$ is the action predicted by strategy $K$ in response to history $\mu(t)$ and $n_{K}(t)$ is the number of agents using strategy $K$ at time $t$. $K$ labels the $K$'th highest scoring strategy while ${\overline K}=2P+1-K$ labels the anti-correlated strategy. In the non-networked MG at low $m$, $D(t)$ exhibits large, crowd-driven fluctuations while $\mu(t)$ follows a quasi-deterministic Eulerian Trail in which all histories $\{\mu(t)\}$ are visited equally \cite{Eulerian}. Hence, the time-averages  $\langle n_{-1}(t)\rangle=\langle n_{+1}(t)\rangle$ yielding $\langle D(t)\rangle=0$ which is the optimal value for $D(t)$. We continue this focus on small $m$ here, since we are interested in the effect of $q$ on these crowd-driven fluctuations. We will assume that the combined effect of averaging over $t$ for a given $\Psi$ (where $\Psi$ is a given realization of the initial strategy allocation matrix \cite{NFJ1}), {\em and} averaging over $\Psi$, will have the same effect as averaging over all histories. This is true for the non-networked MG, and produces a mean
$D(t)$ of zero. Hence the fluctuation (i.e. standard deviation) of the
excess demand,  $\sigma_{D}$, is given by:  
{\small 
\begin{equation}
\sigma^{2}_{D} =   
\left\langle \sum_{K=1}^{P}\left(
n_{K}(t)-n_{\overline{K}}(t)\right) ^{2}\right\rangle _{t, \Psi}
\approx \sum_{K=1}^{P}\left(
n^{\mathrm{mean}}_{K}-n^{\mathrm{mean}}_{\overline{K}}\right)^{2} \ .
\label{var}
\end{equation}}

\noindent We have used the orthogonality properties of the vectors with elements $a_{K}$ where $K=1,2,\dots,2P$ \cite{NFJ1}. Since $n_{K}(t)$ will generally fluctuate around some mean value $n^{\mathrm{mean}}_{K}$, we have also written $n_{K}(t)=n^{\mathrm{mean}}_{K}+\epsilon_K(t)$ and assumed that the fluctuation terms $\{\epsilon_K(t)\}$ are uncorrelated stochastic processes.  In Eq. (2),

\begin{equation}  
n^{\mathrm{mean}}_{K} = n_{K} + {n}_{\rightarrow K} - {n}_{K \rightarrow} - n_{K}^{\mathrm{conn}}
\label{nNet}
\end{equation}

\noindent and similarly for $n^{\mathrm{mean}}_{\overline{K}}$, where:

\noindent (1) $\ n_{K}$ is the mean number of agents whose own best strategy is actually the $K$'th highest scoring strategy in the game \cite{NFJ1}:

%%%%%%%%%% n_{K} %%%%%%%%%%%%%
\begin{equation} n_{K} =  N\Big[ \bigg(1-\frac{K-1}{2^{m+1}}\bigg)^{S}
-\bigg(1-\frac{K}{2^{m+1}}\bigg)^{S} \; \Big]\ \ .
\end{equation}
%%%%%%%%%%%%%%%%%%%%%%%%%%%%%%

\noindent (2) $\ {n}_{\rightarrow K}$ is the mean number of agents who only
possess strategies worse (i.e. lower scoring) than $K$, but who will use strategy $K$ due to connections they have to one or more agents who each possess strategy $K$ but no better:

%%%%%%%%%% n_{-->K} %%%%%%%%%%%%%
\begin{equation} {n}_{\rightarrow K} = \Big(1-q\Big)\cdot n^{q}_{\rightarrow K} + q\cdot n^{q}_{\rightarrow \overline{K}}
\label{inNet}
\end{equation}
%%%%%%%%%%%%%%%%%%%%%%%%%%%%%%%%%

\noindent where 

%%%%%% 1st term of n_{-->K}%%%%%% 
{\small
\begin{equation} n^q_{\rightarrow K} = \Big[\sum_{J>K}n_{J}
\Big]\Big[\bigg(1-p\bigg)^{\sum_{G<K}n_{G}}\Big]\Big[1-\bigg(1-p\bigg)^{n_{K}}\Big]
\label{in}
\end{equation} }
%%%%%%%%%%%%%%%%%%%%%%%%%%%%%%%%%

\noindent with $n^q_{\rightarrow \overline{K}}$ being obtained from Eq. (\ref{in}) by setting $K\rightarrow \overline{K}=2P+1-K$. 

\noindent (3) $\ {n}_{K \rightarrow}$ is the mean number of agents who possess strategy $K$, but who will nevertheless use a strategy better than $K$ due to connections:  

%%%%%%%%%% n_{K-->} %%%%%%%%%%%%%
\begin{equation} {n}_{K \rightarrow} = \Big(1-q\Big)\cdot n^q_{K \rightarrow} +
q\cdot n^q_{\overline{K}\rightarrow}
\label{nOut}
\end{equation}
%%%%%%%%%%%%%%%%%%%%%%%%%%%%%%%%%

\noindent where

%%%%%% 1st term of n_{K-->}%%%%%%
\begin{equation} n^q_{K \rightarrow} = n_{K}\Big[\
1-\bigg(1-p\bigg)^{\sum_{G<K}n_{G}}\ \ \Big]
\label{out}
\end{equation}
%%%%%%%%%%%%%%%%%%%%%%%%%%%%%%%%%

\noindent and similarly for $n^q_{\overline{K}\rightarrow}$.

\noindent (4) $\ n_{K}^{\mathrm{conn}}$ accounts for the situation in which an agent is connected to other agents with the same highest scoring strategy $K$ as him. $q$ therefore gives the probability that this agent will take the opposite action to strategy $K$:

%%%%%%%%%% n_{K}^{corr} %%%%%%%%% 
{\small
\begin{equation} n_{K}^{\mathrm{conn}} = q\cdot n_{K}\bigg[ 1
-\Big(1-p\Big)^{n_{K}}\bigg] -  q\cdot n_{\overline{K}}\bigg[
1-\Big(1-p\Big)^{n_{\overline{K}}} \ \bigg]\ \ .
\label{nCorr}
\end{equation}}
%%%%%%%%%%%%%%%%%%%%%%%%%%%%%%%%%

Figure \ref{figure1} compares the numerical and analytical results for $\sigma _{D}$, which is the standard deviation in excess demand. The agreement is remarkable given the complexity of $\sigma _{D}$ as a function of $p$ and $q$. As the `noise' level $q$ increases, the system undergoes a change in regime at a critical connectivity $p$ defined by the critical boundary $C_{\rm crit}(q,p)$. Moving across $C_{\rm crit}(q,p)$, the symmetry in the global outcome string is spontaneously broken in a manner reminiscent of a phase transition. Specifically, the global outcome series changes from the low-$q$ phase where it resembles the period-4 Eulerian  Trail $\dots 00110011 \dots$,  to a high-$q$ phase where it comprises {\em two} distinct branches (see Fig. \ref{figure1}(a)). Figure \ref{figure2} shows individual runs near the critical noise threshold. The higher branch
corresponds to a period-2 global outcome series $\dots 1010\dots$  which is {\em antiferromagnetic} if we denote 0 (1) as a {\em spatial} spin up (down) as opposed to a {\em temporal} outcome. The lower branch corresponds to the period-1 series of `frozen' outcomes $\dots 0000\dots$ or $\dots 1111\dots$, i.e. {\em ferromagnetic}. In this high-$q$ phase, the system will `choose' one of these two global outcome branches spontaneously, as a result of the type and number of links each agent has.  This symmetry-breaking of the global outcome series along the channel of minimum fluctuation in Fig. \ref{figure1}, $C_{\rm crit}(q,p)$, originates in the internal coupling between the history dynamics, the strategy space and the individual agent networks. Because of the initial strategy allocation and connections, many agents will have an in-built bias towards one of the two possible actions and hence act in a deterministic or `decided' way at a given timestep. However, there exist a few `undecided' agents who need to toss an unbiased coin to decide between the equally balanced signals they gather from their local network. It is the fluctuations of these few `undecided' agents who then push the system onto a particular branch.

We now discuss two technical details. First, there are many ties in strategy scores at very small $m$, and hence many tie-break coin-tosses.
This means that the fluctuation terms $\{\epsilon_K(t)\}$ can no longer be ignored. The $m=1$ surface in Fig. \ref{figure1}(b) was therefore produced by averaging over the $2.2^m=4$ timesteps in the Eulerian Trail \cite{Eulerian}. In other words, the double-average in Eq. (\ref{var}) was evaluated over the $2.2^m=4$ timesteps in the Eulerian Trail. When a tie-break between the strategies $K=1$ and $K=2$ arises at one of the four timesteps, one replaces $n_{K=1}^{\mathrm{mean}}$ and $n_{K=2}^{\mathrm{mean}}$ by
$\frac{1}{2}(n_{K=1}^{\mathrm{mean}}+n_{K=2}^{\mathrm{mean}})$ at that timestep. Likewise, for tie-breaks between any other $K$ and $K'$. In this way, the average over the Eulerian Trail is easily evaluated analytically. As $m$ increases, there are more timesteps over which one must average (i.e. $2P=2.2^m$ timesteps). However, since ties also become less frequent as $m$ increases, one can simply ignore them without significant loss of accuracy (see Ref. \cite{NFJ1} in which good agreement is obtained for the non-networked MG for a wide range of $m$ values without considering ties). 
Second, the theory has assumed the non-networked MG result that the dynamics follow the Eulerian Trail. Only one branch therefore emerges in Fig. \ref{figure1}(b) at high $q$, appearing like some effective average over the global output series for all branches in Fig. \ref{figure1}(a). If instead one uses knowledge of the actual global output series for these separate branches (i.e. antiferromagnetic or ferromagnetic), then results even closer to Fig. \ref{figure1}(a) can be obtained.  This is illustrated at one particular $p$ by the dotted line in Fig. \ref{figure1}(b).

Figure \ref{figure3} provides a contour plot of $\sigma _{D}$ around the minimum.  The black contour, centered around the critical curve  $C_{\rm crit}(q,p)$, effectively separates the two different regimes of behavior. The low $\sigma _{D}$ values around $C_{\rm crit}(q,p)$ can be easily understood using the physical picture provided by the Crowd-Anticrowd theory: the stochasticity induced by $q$ (i.e. noise) breaks up the size of the Crowds using a given strategy $K$, while simultaneously
increasing the size of the Anticrowds using the opposite strategy  $\overline K$. It is remarkable that a linear increase in the `noise' $q$ gives rise to such a non-linear variation in $\sigma _{D}$. Using the
analytic expressions in this paper, an equation for $C_{\rm crit}(q,p)$ can be obtained -- however we do not include it because it is cumbersome. As noted above, the theory neglects a full treatment of the dynamical fluctuations around $n_K^{\rm mean}$. Hence, the theory overestimates the Crowd-Anticrowd cancellation arising in Eq. (\ref{var}) and thus slightly underestimates $\sigma _{D}$ in the neighborhood of $C_{\rm crit}(q,p)$ (compare Figs. \ref{figure3}(a) and \ref{figure3}(b)). As $p$ increases, $C_{\rm crit}(q,p)$ becomes less dependent on the connectivity $p$ since
more and more agents join the same network cluster. For $p \gtrsim 0.05$ the system passes the percolation threshold and hence is dominated by a giant, common cluster. 

Finally, we note that if instead of introducing errors into the {\em local}
transmission of information, as in the present work, one induces noise at the {\em global} level via errors in the global outcome series, the
results for the excess demand are reasonably similar to those presented here. Our results raise the interesting possibility whereby imperfect information transmission could be induced at the local and/or global level in order to achieve a desired change in the macroscopic fluctuations within 
biological, informational or socio-economic systems.   

We are grateful to Sean Gourley for assistance with this work. PMH acknowledges support from the Research Grants Council of the Hong Kong SAR Government (grant CUHK4241/01P).


\begin{thebibliography}{99}
\bibitem{qNote} Defining $q=(1/2)[1-{\rm tanh}(1/T)]$ \cite{TMG}, one could
choose to convert the parameter $q$ into a `physical' temperature $T$ which
creates disorder in the information carried by the `bonds' (as opposed to the `sites' in the TMG \cite{TMG}). However, since $T$ must be positive to be physical, this transformation would restrict $q$ to the range $0\leq q \leq 1/2$, thereby discarding the interesting high $q$ behaviour.
\end{thebibliography}
\end{document}